\documentclass[prd,aps,reprint,showpacs,superscriptaddress]{revtex4-1}
\usepackage{amsmath,amssymb}
\usepackage{graphicx}

\newcommand{\be}{\begin{equation}}
\newcommand{\ee}{\end{equation}}

\usepackage{color}

\newcommand{\pt}{\tilde{p}}
\newcommand{\rhot}{\tilde{\rho}}  

\begin{document}

\title{Bound states of the $\phi^4$ model via the Non-Perturbative Renormalization Group}

\author{F. Rose}
\affiliation{LPTMC, CNRS-UMR 7600, Universit\'e Pierre et Marie Curie, 75252 Paris, France}
\author{F. Benitez}
\affiliation{Max Plank Institute for Solid State Research, Heisenbergstr. 1, 70569, Stuttgart, Germany}
\affiliation{Physikalisches Institut, Universit\"at Bern, Sidlerstr. 5, 3012 Bern, Switzerland}
\author{F. L\'eonard}
\affiliation{LPTMC, CNRS-UMR 7600, Universit\'e Pierre et Marie Curie, 75252 Paris, France}
\author{B. Delamotte}
\affiliation{LPTMC, CNRS-UMR 7600, Universit\'e Pierre et Marie Curie, 75252 Paris, France}

\begin{abstract}
Using the nonperturbative renormalization group, we study the existence of bound states in the symmetry-broken phase 
of the scalar $\phi^4$ theory in all dimensions between two and four and as a function of the temperature.
The accurate description of the momentum dependence of the two-point function, required to get the spectrum of the theory,
is provided by means of the Blaizot--M\'endez-Galain--Wschebor
approximation scheme. We confirm the existence of a bound state in dimension three, with a mass
within $1\%$ of previous Monte-Carlo and numerical diagonalization values. 
\end{abstract}


\date{\today}

\maketitle

\section{Introduction}
\label{intro}

The low-energy physics of a many-body system is governed by the first excitations above its ground state. Often, 
these excitations are not directly given by the (dressed) elementary constituents of the system, but are instead complicated objects.
Bound states represent an important class of such excitations. 
Superconductivity is probably the best known  
example \cite{degennes99}, although not the only one of experimental interest: low dimensional quantum systems, such as 1d cuprate 
ladder materials, or the 2d magnetic system SrCu$_2$(BO$_3$)$_2$, are known to also present bound states, whose modes appear in their 
measured spectra \cite{uhrig96,trebst00,knetter01,zheng01,windt01}. Beyond the realm of condensed matter, bound states are also very important in the theory of nuclear forces \cite{ZINN,cohen11}, as well as in quantum chemistry \cite{levine09}.

Of particular interest are bound states emerging in strongly correlated systems,
where they are less understood and difficult to characterize using standard perturbative techniques.
Examples of such systems arise in the theory of the strong nuclear interactions, QCD \cite{jaffe77,shifman79}, as well as in the spectrum of quantum excitations in strongly correlated electron systems \cite{vidal02,vidal00,capponi07,sachdev}. There is no need to stress that the study of this class of subjects is ripe for the development of new 
approximation methods, capable of dealing with the complexities that emanate from strong correlations.
In this regard, Renormalization Group methods have already shown to be extremely well adapted to this endeavor, given their focus on scale dependent properties, and their proven abilities to deal with strong correlations.

The scalar $\phi^4$ field theory probably represents the simplest example of a strongly correlated system, exhibiting long-range order and a diverging correlation length at the critical point. Near this point, it belongs to the Ising universality class, and a lot of effort has 
been put into the understanding of its properties by a myriad of methods, 
including Monte-Carlo simulations, high order perturbative expansion \cite{Guida98,hasen10,campos06} and the conformal bootstrap \cite{rychkov12,gliozzi14,rychkov14,kos14,rychkov14b}. 
The complex yet manageable behavior of $\phi^4$ theory is thus ideally suited for purposes of benchmarking new  approximation methods  \cite{BMWpre,BMWlong}, as well as for revisiting a subject of intrinsic interest.

In two spatial dimensions, or, equivalently, in the 1d quantum case at zero temperature \cite{sachdev}, 
the integrability of the Ising model, and thus of its universality class close to criticality, 
allows for the complete determination of the bound state spectrum \cite{zamolo1,zamolo2}. These exact 
results stem from the conformal invariance of the theory at criticality. In the language of the classical two-dimensional model, this bound state exists in the presence of a small magnetic field, and when the temperature is exactly the critical temperature. The observed ratio between the mass of the first bound state and of the elementary excitation is $m_1/m_0 = (1+\sqrt{5})/2$ (golden ratio), which has been experimentally confirmed 
\cite{exp} in the quasi-1d quantum Ising ferromagnet CoNb$_2$O$_6$, by means of inelastic neutron scattering. Six other bound states are known to exist in this case with only two that are below the mass of the threshold $2m_0$ of the multi-particle continuum.

In the case of the three dimensional $\phi^4$ model, the presence of a bound state in the symmetry broken phase is by now a well-established fact.
A classical argument at $T=0$ shows the existence of a bound state for the Ising model \cite{hasen97}, although this is of course a non-universal result. 
For the $\phi^4$ theory, a bound state with a mass ratio $M/m$ around $1.8$ was first detected by Monte Carlo simulations \cite{hasen97,hasen99} of the Ising model at temperatures lower than $T_c$, but still within the scaling regime, and thus expected to be universal. This prompted the use of resummed perturbative calculations by means of a Bethe-Salpether equation, where the leading order yields results compatible with Monte Carlo values \cite{hasen00,hasen02}. However, the next-to-leading order result leads to an unphysical conclusion $M/m < 0$, indicating a strongly non-perturbative behavior for this quantity. Alternatively, this bound state can also be detected studying the quantum 2d Ising model at zero temperature, as was performed using high-order perturbative continuous unitary transformations \cite{vidal10}. More recently, Nishiyama \cite{ponja} found $M/m = 1.84(1)$ using numerical diagonalization methods.

For the three dimensional $\phi^4$ model then, the observed bound state appears to be outside of the standard perturbative Renormalization Group regime. It is interesting however to see whether other, non perturbative, methods based on the RG are able to detect its presence. One of the main difficulties is that a non-trivial 
(and in particular, non-analytic) description of the momentum dependence of the correlation 
functions of the system is needed to detect bound states, as is discussed below. 

Some years ago, a new approximation scheme of the Non Perturbative Renormalization Group 
(NPRG, also known sometimes as the Functional or Exact RG), the Blaizot--M\'endez-Galain--Wschebor (BMW) approximation, has been shown to give accurate momentum 
dependent results for scalar field theories \cite{BMW,BMWlong}. In this work, we use this approximation to compute the $\phi^4$ bound state mass within the NPRG, for spatial dimensions between $d=2$ and $d=4$. This not only shows the strength of this multi-purpose method, but also allows us to study in a novel way the temperature-dependence of this bound state, even in the non-universal region of the model.

This paper is organized as follows: in section \ref{2point}, we discuss how to check for bound states 
using the momentum dependence of the two-point correlation function of the system, and section \ref{BMWpourlesnuls} briefly presents the approximation scheme used for obtaining the full momentum dependence of this function. 
Section \ref{num} discusses the numerical implementation, as well as the numerical analytic continuation 
procedure, before presenting our main results in section \ref{results}. Finally, we present our conclusions in section \ref{conclusion}.

\section{Signature of a bound state in the spectral function} 
\label{2point}

For concreteness, we use the language of classical equilibrium statistical mechanics, but the case of $d-1$ 
quantum statistical systems at zero temperature corresponds to a trivial renaming of the fields. 
The microscopic Euclidean action of the model is written in the well-known Ginzburg-Landau form \cite{ZINN}
\be
S[\varphi] = \int d^dx \left\{ \frac 1 2 \big(\nabla \varphi(x) \big)^2 + \frac{r_0}{2} \varphi^2(x) + \frac {u_0}{4!} \varphi^4(x) \right\}.
\ee

When performing  Monte Carlo simulations of this system on a lattice, bound states can be most easily 
detected by studying the spatial behavior of the two-point connected correlation function. 
In the symmetry broken phase, one expects correlations decaying exponentially with distance
\begin{equation}
\label{2pointreal}
\langle\varphi(x) \varphi(0) \rangle_c \underset{x \to \infty}{\sim} Ae^{-mx},
\end{equation}
with $m=\xi^{-1}$ the inverse correlation length, usually termed the ``mass'' in analogy with Quantum Field Theory.
For a theory with a non-trivial spectrum,  sub-leading exponentials are expected as well:
\begin{equation}
\label{2pointreal2}
 \langle\varphi(x) \varphi(0) \rangle_c \underset{x \to \infty}{\sim} A_0e^{-m x} + A_1e^{-Mx} + \ldots
\end{equation}
which are associated with bound states of the theory, in that they give the sub-leading  correlation lengths. 
While this is the standard technique for finding bound states when using  simulations \cite{hasen97}, 
one can alternatively study the momentum-dependent spectral function, defined by the Fourier transform
\begin{equation}
 G(p) = \int d^dx \,\langle\varphi(x) \varphi(0) \rangle_c \,e^{-ipx}.
\end{equation}
The presence of sub-leading exponential decay terms can also be  seen in the analytic 
continuation of $G(p)$ to complex values of $p$. Indeed, $G(p)$
behaves in the infrared limit $p\to0$ as 
\begin{equation}
 G(p)\underset{p \to 0}{\sim} \frac{A'_0}{p^2+m^2}+\frac{A'_1}{p^2+M^2}+\cdots
\end{equation}
This implies that the function $G(\omega=ip)$  has poles 
at the values of the masses of the system, with the first mass associated with the correlation length and the 
following with bound states or possible many-particle states.

It can be shown \cite{hasen00,hasen02} that at any (finite) order in a perturbative expansion around a free theory, 
the ratio of the correlation length to any other length scale must be an integer, forbidding thus the description of bound states. 
This issue can be partially solved by performing infinite-order resummations 
in some particular channel, but then this expansion seems to be badly behaved \cite{hasen00}.

Not being able to see a non-integer $M/m$ ratio is a problem shared by the simplest approximation schemes within the NPRG, such 
as the well-known Local Potential Approximation (LPA), or its higher order generalization dubbed 
the Derivative Expansion (DE) \cite{Berges00,Morris94c}. 
 This approximation  amounts to a small momentum expansion of the (vertex) correlation functions.  
While it has proven to be very accurate in the low momentum regime, e.g. for the determination of critical 
exponents \cite{Berges00,delamotte03,canet03a,canet04a,canet04b,canet05a}, it is not reliable for finite momentum properties and is therefore unable to describe bound states.

As already mentioned in the Introduction, the BMW scheme \cite{BMW} takes into account the full, non-trivial momentum dependency of the correlation functions. This method has been successfully 
applied to O$(N)$ scalar field models \cite{BMWlong,BMWONold}, showing excellent results for universal properties such 
as critical exponents and momentum-dependent scaling functions. The BMW method has also found applications beyond 
the confines of equilibrium statistical mechanics \cite{kpz,adam14,felix}, showing its flexibility 
to deal with highly non-trivial momentum dependent quantities. It thus seems very natural to apply this scheme 
to the problem of bound states. We present the BMW scheme and its application to $\phi^4$ theory in the following section.

\section{Non-trivial momentum dependence within the NPRG: the BMW approximation}
\label{BMWpourlesnuls}

We start with a brief outline of the NPRG formalism for the case of a scalar field theory \cite{Wetterich92,Ellwanger93,Tetradis94,Morris94b,Berges00}. 
The NPRG strategy  is to build a family of theories indexed by a 
momentum scale $k$,   
such that fluctuations are smoothly taken into account as $k$ is lowered 
from the  microscopic scale $\Lambda$ (e.g. the inverse lattice spacing) down to 0. 
In practice, this is achieved by adding  to the original  
Euclidean action $S$ a ``mass-like'' term of the form   
$\Delta S_k[\varphi]= \frac{1}{2} \int_q\: R_k(q^2)\varphi(q)\varphi(-q)$ (here $\int_q\equiv \int \frac {d^dq}{(2\pi)^d}$). 
The cut-off function $R_k(q^2)$ is chosen such that $R_k(q^2)\sim k^2$ for $q\lesssim k$, which 
effectively suppresses the modes $\varphi(q\lesssim  k)$, and such that 
it (almost) vanishes for $q\gtrsim k$, leaving 
the modes $\varphi(q\gtrsim k)$ unaffected. 
One then defines a  scale-dependent partition function \be
{\cal Z}_k[J] = \int\! {\cal D}\varphi\; 
e^{-S[\varphi]- \Delta S_k[\varphi] +\int\! J\varphi} \;, 
\label{zk}
\ee
and a scale-dependent 
effective action $\Gamma_k[\phi]$ through a (modified) Legendre transform \cite{Berges00},
\be
\Gamma_k[\phi] +\log {\cal Z}_k[J] = 
\int_q J(q) \phi(-q) -\frac{1}{2} \int_q R_k(q^2) \phi(q)\phi(-q)\,, 
\label{legendre}
\ee
 with $\phi=\delta \ln{\cal Z}_k/\delta J$ the mean value of the field. 
 The variation of the effective action $\Gamma_k[\phi]$ as $k$ changes is given by the Wetterich
equation\cite{Wetterich92}:
\be
\partial_k\Gamma_k[\phi] = \frac{1}{2} \int_q \partial_k{R}_k(q^2)\, G_k[q,\phi] \;,
\label{rgexact}
\ee
where $G_k[q,\phi]=(\Gamma^{(2)}_k[q,\phi]+R_k(q^2))^{-1}$, 
and $\Gamma^{(2)}_k[q,\phi]$ is the second functional derivative of 
$\Gamma_k[\phi]$ w.r.t. $\phi$. 

With the definitions above, it is easy to show that for $k=\Lambda$, all  fluctuations are frozen by the 
$\Delta S_k$ term and thus $\Gamma_{k=\Lambda}[\phi]=S[\phi]$. This is the initial condition of the flow equation
(\ref{rgexact}). On the other hand, when $k=0$, $\Delta S_{k=0}\equiv 0$ because $R_{k=0}(q^2)$ vanishes identically 
and $\Gamma_{k=0}[\phi]$ is the Gibbs free energy of the original model that we aim to compute.

Differentiating $s$ times Eq.~(\ref{rgexact})  with respect to the field $\phi(q)$
yields the flow equation for the vertex function 
$\Gamma^{(s)}_k[q_1,\dots,q_s;\phi]$. Thus, for instance,  the flow equation for $\Gamma^{(2)}$ 
evaluated in a constant field configuration $\phi$ reads:
\begin{equation}
\begin{array}{l}
{\displaystyle\partial_k\Gamma^{(2)}_k(p,\phi)= 
\int_q \partial_k{R}_k(q^2) G_k^2(q)\left[
\Gamma^{(3)}_k(p,\!-p\!-\!q,q)\times\right. }\\ 
\;\;
\left. G_k(p\!+\!q)\Gamma^{(3)}_k(\!-p,p\!+\!q,\!-q)
\!-\! \frac{1}{2}\Gamma^{(4)}_k(p,\!-p,q,\!-q)\right].
\end{array}
\label{rggamma2}
\end{equation}
 (Here we have omitted the $\phi$ dependence of the functions $G_k$ and $\Gamma_k^{(n)}$ in the right hand side to alleviate the notation).
Note that the flow equation for $\Gamma^{(s)}_k(q_1,\dots,q_s,\phi)$ involves 
$\Gamma^{(s+1)}_k$ and $\Gamma^{(s+2)}_k$, leading to an infinite hierarchy of coupled equations. 

The flow equations (\ref{rgexact}) and (\ref{rggamma2}) are exact, 
but solving them requires in general approximations. It is precisely one of the virtues of the NPRG to 
suggest approximation schemes that are not easily derived in other, more conventional approaches.  
In particular, one can develop approximation schemes  for the effective action itself, that is, which apply to 
the entire set of correlation functions. The  BMW approximation\cite{BMW} is such a scheme. It relies on two observations. 
First, the presence of the cut-off function $R_k(q^2)$  guarantees
the smoothness of the $\Gamma^{(s)}_k$'s for $k>0$ and limits  the internal momentum $q$ in equations such as (\ref{rggamma2})
to $q\lesssim k$. In line with this observation, one neglects the $q$-dependence of the vertex functions in the r.h.s. 
of the flow equations (e.g. in $\Gamma^{(3)}$  and $\Gamma^{(4)}$  in Eq.~(\ref{rggamma2})),  
while keeping the full dependence on the external momenta $p_i$.  
The second observation is that, for uniform fields,
$\Gamma^{(s+1)}_k(p_1,\dots,p_s,0,\phi)=
\partial_\phi \Gamma^{(s)}_k(p_1,\dots,p_s,\phi)$, 
which enables one to close the hierarchy of NPRG equations. 

At the leading order of the BMW scheme, one keeps the non trivial momentum dependence of the 
two-point function and implements the approximations above  on Eq.~(\ref{rggamma2}), which
becomes:
\begin{equation}
k\partial_k\Gamma^{(2)}_k(p,\phi)  =  J_3(p,\phi) \left({\partial_\phi\Gamma^{(2)}_k}\right)^2
  \!\! -\! \frac{1}{2} J_2(0,\phi)\, \partial_\phi^2\Gamma^{(2)}_k
\label{BMW}
\end{equation}
with 
\begin{equation}
J_n(p,\phi) \! \equiv \! \int_q k\partial_k{R}_k(q^2) \, G_k^{n-1}(q,\phi) G_k(p\!+\!q,\phi)\;.
\label{integraleJ}
\end{equation}
The approximation can be systematically improved: The order $s$  consists in keeping the full momentum 
dependence of $\Gamma^{(2)}_k,\dots, \Gamma^{(s)}_k$ and truncating that of  
$\Gamma^{(s+1)}_k$ and $ \Gamma^{(s+2)}_k$ along the same lines as those leading to
Eq.~(\ref{BMW}) for the case $s=2$, with a corresponding increase in the numerical complexity.

In order to treat efficiently  the low (including  zero) momentum sector,
we work with dimensionless and renormalized quantities.
Thus, we measure all momenta  in units of $k$: 
$\tilde{p}=p/k$. We also rescale $\rho\equiv \frac{1}{2}\phi^2$  according to 
$\tilde{\rho}=k^{2-d} Z_k K_d^{-1}\,\rho$ (with the constant $K_d=(2\pi)^{-d} S_d/d$, and $S_d$ 
being the volume of the unit sphere), and set  $\tilde\Gamma^{(2)}_k(\tilde{p},\tilde{\rho})=
k^{-2}Z_k^{-1}\Gamma^{(2)}_k(p,\rho)$. 
The running anomalous dimension $\eta_k$ is defined by 
$k\,\partial_k Z_k= -\eta_k Z_k$, so that at criticality $\eta_{k\to0}\to\eta$, with $\eta$ the anomalous dimension. Thus,
at a fixed point $Z_k\sim k^{-\eta}$. The absolute normalization of  $Z_k$ is fixed by 
choosing a point $(\tilde{p}^*,\tilde{\rho}^*)$ where 
$\partial_{\tilde{p}^2}\tilde{\Gamma}^{(2)}_k\vert_{\tilde{p}^*,\tilde{\rho}^*}=1$.
 Here, we have chosen $\pt^*=0$ and $\rhot^*=\rhot_{0,k}$, 
where $\rhot_{0,k}$ is the $k$-dependent running minimum of the potential.
Then, the flow equation of 
$\tilde\Gamma^{(2)}_k(\tilde{p},\tilde{\rho})$ follows trivially from 
Eq.(\ref{BMW}). 

It is actually more accurate to disentangle the potential part of $\tilde\Gamma^{(2)}_k(\tilde{p},\tilde{\rho})$
from the momentum part and to solve independently the flows of these two quantities \cite{BMWlong}. We thus solve two equations: one for
$\tilde{Y}_k(\tilde{p},\tilde{\rho}) \equiv \tilde{p}^{-2} 
[\tilde{\Gamma}^{(2)}_k(\tilde{p},\tilde{\rho})-
\tilde{\Gamma}^{(2)}_k(0,\tilde{\rho})]-1$ and one for
the derivative of the dimensionless effective potential
$\tilde{W}_k(\tilde\rho)=\partial_{\tilde \rho} \tilde{V}_k(\tilde \rho)$, with $\tilde{V}_k(\tilde \rho)=K_d^{-1}k^{-d}V_k(\rho)$.
 Note that 
 $\tilde{\Gamma}^{(2)}_k(0,\tilde{\rho})=
\tilde{W}_k(\tilde\rho) + 2 \tilde\rho\,\tilde{W}_k'(\tilde\rho) $.  
Here and below, primes denote derivative w.r.t.  $\tilde\rho$, or, in the case of dimensionful variables, w.r.t. $\rho$.
These two equations read  (dropping the  $k$ index and the $\tilde{\rho}$ and $\tilde{p}$ dependences to alleviate the notation):
\begin{align}
\partial_t \tilde{Y} ={}&  \eta_k (1+\tilde{Y})+\tilde{p}\, \partial_{\tilde{p}} \tilde{Y} -(2-d-\eta_k)\tilde{\rho}\,\tilde{Y}'\nonumber\\ 
& + 2{\tilde{\rho}}\,\tilde{p}^{-2}\left[(\tilde{p}^2\,\tilde{Y}'\!+\!\tilde\lambda_k)^2 \tilde{J}_3(\tilde{p},\tilde{\rho})
-\tilde\lambda_k^2 \tilde{J}_3(0,\tilde{\rho})\right] \nonumber\\
& - \tilde{J}_2(0,\tilde{\rho})(\tilde{Y}'/2+\tilde{\rho}\,\tilde{Y}''),\label{eqY} \\
\partial_t \tilde{W} ={}&  (\eta_k \!-\! 2) \tilde{W} +(d \!-\! 2 \!+\! \eta_k)\tilde{\rho}\,\tilde{W}'+ \frac{1}{2} \tilde{J}_1'(0,\tilde{\rho}). \label{eqW}
\end{align}
Here, the renormalization ``time'' $t$ is defined by $t=\log k/\Lambda$, $\partial_t=k\partial_k$,
$\tilde{J}_n(\tilde{p},\tilde{\rho})= K_d^{-1}Z_k^{n-1}k^{2n-d-2}{J}_n({p},{\rho})$ 
and $\tilde{\lambda}_k(\tilde\rho)=
3\tilde{W}_k'(\tilde\rho)+2\tilde\rho\,\tilde{W}_k''(\tilde\rho)$. The running anomalous dimension
$\eta_k$ is obtained by setting 
$\tilde{Y}_k(\tilde{p}^*,\tilde{\rho}^*)=0$ in Eq.(\ref{eqY}) and taking a time derivative, 
noting that $ \partial_t \rhot_{0,k}=- \partial_t  \tilde{W}_k(\tilde\rho)|_{\rhot_{0,k}}/\tilde{W}'_k(\rhot_{0,k})$.

At the end of the numerical flow, the two-point vertex function $\Gamma^{(2)}(p,\rho)$ can be reconstructed from $\tilde{Y}_k$ and $\tilde{W}_k$:
\[
 \Gamma^{(2)}(p,\rho)=\lim_{k\to0}Z_k k^2 \bigg[ \tilde{p}^2\Big(1+\tilde{Y}_k(\tilde p,\tilde \rho)\Big)+\tilde{W}_k(\tilde \rho)+2\tilde\rho \tilde{W}_k'(\tilde \rho) \bigg].
\]

The flow equations (\ref{eqY}) and (\ref{eqW}) can be solved using standard numerical techniques, 
see section \ref{num}. Crucially for our purposes, the obtained $k \to 0$ values for 
the $\Gamma^{(2)}(p)=G(p)^{-1}$ function must then be extended to the whole complex plane. 
This is not a completely well-defined mathematical operation, given that we are extending a 
discrete set of numerical values to the whole plane. Nonetheless, this type of extension is often used e.g. 
in systems studied using Quantum Monte Carlo methods \cite{Pade,PadeClean}, and are known to yield useful results. 
Here, we perform this analytic continuation from Pad\'e approximants of our results, see the next section.

 Note that if we neglect non-trivial momentum dependence and set $\tilde{Y}=0$, in the ordered phase $\Gamma^{(2)}(p,\rho=\rho_{0})$ cancels only for $p=\pm i \Delta$,
\begin{equation}
\Delta=\sqrt{\dfrac{2\rho_{0} W_{k=0}'(\rho_0)}{Z_{k=0}}}.
\end{equation}
In this case, $\Delta$ is the mass (the inverse correlation length) and there are no bound states. 
Within the BMW approximation, this is no longer true and bound states can exist. The actual mass $m$ 
of the elementary excitation, that is, the leading correlation length, is close to $\Delta$. 
It is therefore useful to retain $\Delta$ as a relevant energy scale, which needs not to be extracted 
from an analytic continuation.

It should be mentioned that, recently, an alternative formulation of the NPRG capable of dealing 
with the analytic continuation of spectral functions has been proposed in a slightly 
different context \cite{vonsmekal14,vonsmekal14b,vonsmekal14c}. In this approach, the analytic 
continuation is performed at the level of the flow equations of the NPRG, so that one ends up 
with a flow of complex quantities. This is a very promising approach, which has so far only been 
applied together with Derivative Expansion-like approximations, resulting in a less accurate 
momentum description of the theory than our approach. Ideally, these ideas could be implemented together with a 
BMW type of approximation in the near future.

\section{Numerical Procedure}
\label{num}

In this section, we give the key points of the numerical integration of the flow 
equations (\ref{eqY},\ref{eqW}), and then we detail the analytic continuation used to extract 
the pole of the correlation function, where, as shown below, subtle issues may arise.

\subsection{Flow integration}

The integration of the flow equations is based on well-established numerical analysis methods.
The (renormalization) time evolution is done through explicit Euler integration scheme with a time step $dt=-10^{-4}$.
The  momentum dependence of $\tilde{Y}_k(\tilde p,\tilde\rho)$ is studied on the interval $\tilde{p} \in [-\tilde{p}_{\rm max},\tilde{p}_{\rm max}]$ 
with $\tilde{p}_{\rm max}=10$ and we use a 
Chebyshev pseudo-spectral approximation of this function with a variable number of polynomials ranging from 20 to 50.  
The field-dependence of $\tilde{Y}_k(\tilde p,\tilde\rho)$ is obtained by discretizing the $\tilde\rho$-space on a finite and regular grid 
$\tilde{\rho} \in [0,\tilde{\rho}_{\rm max}]$ with $\tilde{\rho}_{\rm max}$ comprised between 10 and 14.
The lattice spacing of the $\tilde{\rho}$ grid  is $d\tilde{\rho} = 0.1$. 

We choose an exponential regulator,
\begin{equation}\label{regulator}
R_k(q^2)= \alpha \dfrac{Z_k q^2}{\exp(q^2/k^2)-1},
\end{equation}
where $\alpha$ is an arbitrary parameter that is varied to study the sensitivity of our results with the choice of regulator.
Since the integrals over the momenta involved in the RG flows, Eq.~(\ref{integraleJ}), are all 
exponentially cut-off by the (derivative of the) regulator, we restrict their range
to $\vert\tilde p\vert \leq 4$. 
These integrals are then computed using a Gauss-Legendre approximation with $40$ points. A good numerical accuracy of the integrals is mandatory to obtain converged results, specially for dimensions $d<3$.

Further difficulties arise when studying the ordered phase. First of all, in this phase, 
$\rho_{0,k}$, which is the minimum of the running potential, goes to a constant value $\rho_{0}>0$ as $k$ goes to zero,
since it is half the square of the spontaneous magnetization. Since $\rhot_{0,k} \sim \rho_{0,k}/Z_kk^{d-2}$, 
this means that $\rhot_{0,k}$ diverges as $k$ goes to zero. In practice, starting close to criticality,
we observe that in the first stage of the flow $\rhot_{0,k}$ evolves towards its fixed point value. Then,
when $k$ is of the order of the inverse of the correlation length $\xi$, it starts diverging which means
that $\rho_{0,k}$ has almost converged to its final value. Hence, when the flow leaves the critical regime, 
we switch to the flow of $\tilde W_k(\rho)$ and $\tilde Y_k(\tilde p,\rho)$ (instead of $\rhot$) while keeping the same number of points on a dimensionful grid in $\rho$.
This allows $\rho_{0,k}$ to remain inside the grid as $k$ goes to zero. 
 
The second difficulty is numerical. 
When $k\to0$, the inner part of the potential, $\rho< \rho_{0}$, becomes
flat because of the convexity of the effective potential $V=V_{k=0}$. The convexity of $V$
is reproduced within the BMW approximation and corresponds to the approach of the pole of
the propagator at vanishing momentum
when $k\to0$. Thus, $G_k(p=0,\rho<\rho_{0,k})=\left [R_k(p=0)+W_k(\rho)\right ]^{-1}$  becomes very large at small $k$ which causes numerical instabilities.
The instabilities arise at small $\rho$ since $W_k(\rho)$ is an increasing function of $\rho$.
Since the physics we are interested in corresponds to $\rho=\rho_0$ (there is neither external magnetic field nor
phase coexistence), we eliminate the source of numerical instabilities by eliminating the small values 
of $\rho$ from the grid, that is, the values for which $W_k(\rho)$ is `too negative'. 
We therefore replace the grid $\rho\in [0,\rho_{\rm max}]$ by a $k$-dependent grid $\rho\in [\rho_{\rm min}(k),\rho_{\rm max}]$
which allows us to continue the flow to smaller values of $k$. 
We expect that this supplementary approximation has a small impact on the final results.
However, a difficulty remains for small $k$: $\rho_{\rm min}(k)$ becomes close to $\rho_{0,k}$ and there are no longer 
enough points in the $\rho$-grid on the left of $\rho_{0,k}$ to compute the derivatives of the potential at this point. The flow
must then be stopped and the smallest value of $k$ we have been able to reach is typically $k_{\rm min}\simeq 0.1\Delta$. 
 
The function $\Gamma^{(2)}(p,\rho_0)$ is finally obtained using the approximation:
\begin{equation}
 \Gamma^{(2)}_{k=0}(p)\simeq \Gamma^{(2)}_{k=p/\pt_{\rm max}}(p),
 \label{approx}
\end{equation}
$k=p/\pt_{\rm max}$ 
 being the smallest value of $k$ for which $p/k$ is still in the 
dimensionless grid $[0,\tilde p_{\rm max}]$.  This approximation is justified by the fact that  $p$ acts as an effective 
infrared cutoff in the flow of $\Gamma^{(2)}_k(p)$ that therefore effectively stops for $k\ll p$. Thus, stopping the flow of $\Gamma^{(2)}_k(p,\rho_{0,k})$ at
$k\ll p$ or at $k=0$ should yield almost the same result. We have checked the validity of the approximation (\ref{approx}) by varying $\pt_{\rm max}$
and observing that it is indeed almost insensitive to $\pt_{\rm max}$ when it is of  order  10.

\subsection{Analytic continuation}

  Let us now detail the Pad\'e approximation procedure used to obtain the spectral function 
 $G(\omega=ip-0^+)=G_{k=0}(i p-0^+,\rho_0)$. First, we compute the propagator $G(p)$ for  
 $N$  momentum values $p_i$ evenly
 spaced in a window $[\omega_{\rm min}, \omega_{\rm max}]$. 
Typically, $N\sim 30-50$, $\omega_{\rm min}\sim  \Delta$ and $\omega_{\rm max} \sim 10 \Delta$. 
We then construct a  $[N-2/N]$ Pad\'e approximant \cite{Pade, PadeClean}  $F$, even in $p$, that satisfies 
$F(p_i)=G(p_i)$ for all $i$. Once $F$ is known, we evaluate ${\rm Im} \left [F(\omega=ip-0^+)\right ]$ 
as an approximation of ${\rm Im}[ G(\omega)]$ that shows peaks where $G(ip)$ has poles. 
 
   \begin{figure}[ht]
 \includegraphics{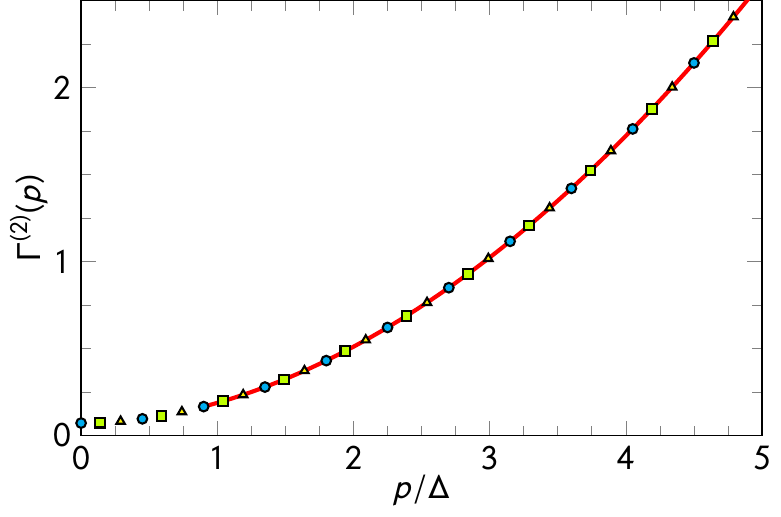}
 \includegraphics{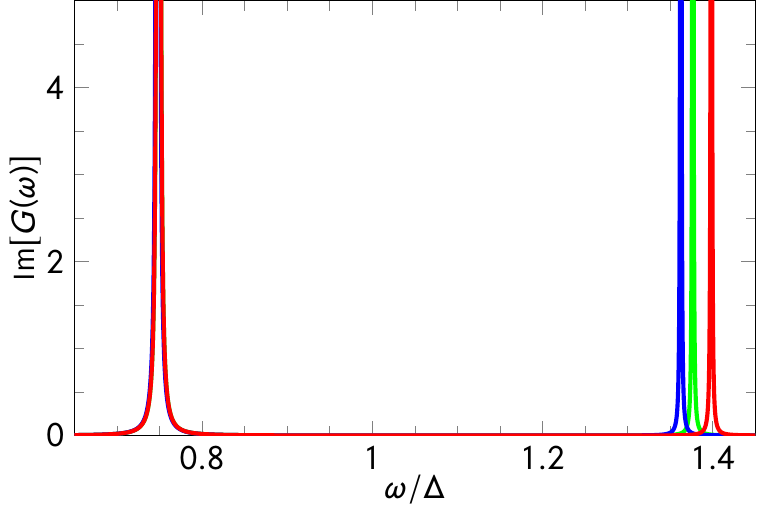}
 \caption{(Color online) Comparison of three different Pad\'e approximants of (top) $\Gamma^{(2)}(p)$, (bottom)
 ${\rm Im}[G(\omega)]$. Top: we show (in arbitrary units) the vertex $\Gamma^{(2)}(p)$ 
 (full line), obtained from the numerical integration of the flow, as well as its fit by three 
 different approximants (symbols). Bottom: the spectral functions ${\rm Im} [G(\omega)]$ (in arbitrary units) obtained from 
 the  analytic continuation of these approximants. The three approximants show two poles, 
 one at a mass $m\simeq 0.75\Delta$ whose position is very stable among the approximants, the other 
 at a mass $M\simeq 1.35-1.4\Delta$ whose position depends slightly on the approximants.}
 \label{fig:padecompared}
 \end{figure}

 To check the validity of this method, we vary the parameters $N$, $\omega_{\rm min}$ and 
 $\omega_{\rm max}$ and compare about 20 different approximants, see Fig. \ref{fig:padecompared}. 
While they all (almost) coincide for real values of $p$, they vary a lot more when analytically continued, 
a signature of the fragility of the Pad\'e procedure with respect to numerical errors. 

All approximants show 
a remarkable agreement for the pole at  $\omega/\Delta$ close to 1, 
corresponding to the mass $m$, that is, the inverse correlation length of the system, 
 see Fig. \ref{fig:padecompared} where all curves are superimposed at this pole. Among the approximants, we eliminate those that exhibit 
unphysical spurious behavior, such 
as an additional pole at an energy $\omega \ll \Delta$, or a splitting of the mass pole 
into two peaks of energy around $m$.  Furthermore, we eliminate approximants which 
present a mass more than $1\%$ different from the others. Depending on the dimension, between one fourth ($d=3$) and
one half ($d=2$) of the Pad\'e approximants are rejected this way.

We observe that all the remaining approximants present a single second pole 
at an energy $M>m$, the value of $M$ varying slightly from approximant to approximant (from $2\%$ in $d=3$ to less than $10\%$ for $d\lesssim 2.6$). Depending on the dimension, we find two possibilities.
 In the first case, $M \gtrsim 2m$, and the pole corresponds to two independent single-particle excitations 
implying that there are no  bound states. In the second case $m<M<2m$ and a bound state exists with mass $M$. 

As an additional check of the accuracy of the analytic continuation, we have verified that 
 the position of the poles varies smoothly with the dimensionality of the system.

\section{Results}
\label{results}

 \begin{figure}[t]
 \includegraphics{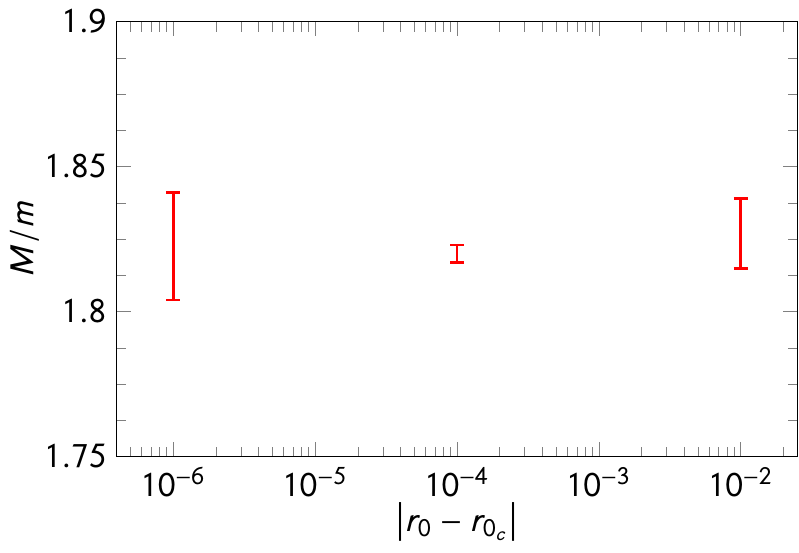}
 \caption{(Color online) Values of the mass ratio $M/m$ for several values of the reduced temperature, 
 corresponding to different values of  $r_0-r_{0_c}$ measuring the distance to criticality. 
 For each temperature, the error bar indicates the extremal possible values obtained from the Pad\'e approximants.}
 \label{fig:massratio}
 \end{figure}
 
 Let us start by discussing the results obtained in $d=3$, where a bound state is clearly present in the broken symmetry phase, and absent in the symmetric phase. The corresponding values of $M/m$ are displayed in Fig. \ref{fig:massratio} as a function of $r_0-r_{0_c}$, where $r_{0_c}$ is the value of the parameter $r_0$ which makes the model critical. For a given value of the reduced temperature (which we identify with $r_0-r_{0_c}$), the value of the ratio $M/m$ varies slightly between different approximants, which is origin of the error bars shown in the figure. To test the accuracy of the method, we have also studied the variation of the results with the parameter $\alpha$ in front of the the regulator function (\ref{regulator}). In all cases this variation turns out to be much smaller than the error bars stemming from the Pad\'e procedure.
 
 Both in the universal regime $r_0 \simeq r_{0_c}$, as well as for larger values of the reduced temperature within the non-universal regime, the ratio does not appear to vary significantly with the reduced temperature. Using a conservative error bar, we find $M/m=1.82(2)$, in agreement with previous results: $1.83(3)$ for Monte Carlo \cite{hasen99}, $1.828(3)$ for the first order approximation of the Bethe-Salpeter equation \cite{hasen02}, $1.84(3)$ for the results of perturbative continuous unitary transformations, and $1.84(1)$ for the most recent and accurate results from numerical diagonalization methods \cite{ponja}. 

 \begin{figure}[t]
 \includegraphics{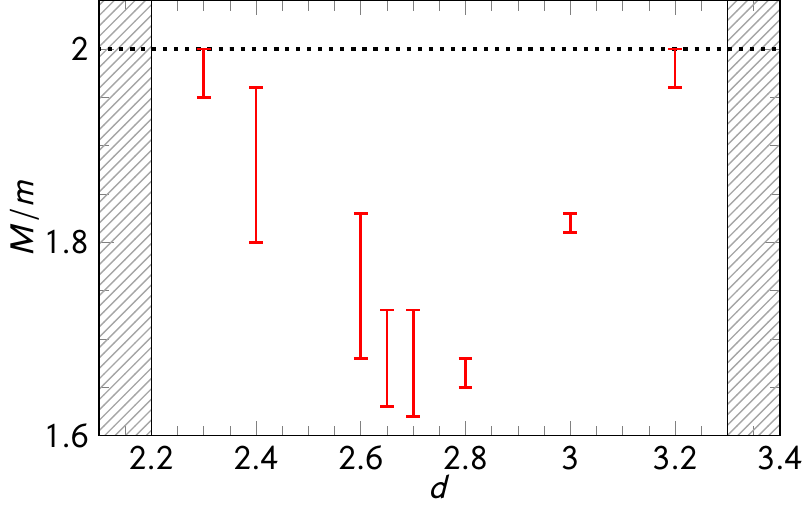}
 \caption{(Color online) Values of the mass ratio $M/m$ in the critical regime for various 
 dimensions. For each dimension, the error bar indicates the extremal possible values obtained 
 from the Pad\'e approximants. The shaded areas denotes the range of dimensions for which it 
 is certain there are no bound states.}
 \label{fig:ratiodim}
 \end{figure}

Next, we study the evolution of value of the $M/m$ ratio at criticality as a function of the 
dimension, for $2\leq d\leq 4$. The ratio is a smooth function of the dimension, as 
shown in Fig. \ref{fig:ratiodim}. It is found that there exists an upper and lower 
dimension, $d_r^-\sim2.2-2.3$ and $d_r^+\sim 3.2-3.3$, such that for $d_r^-<d<d_r^+$ 
there is a bound state, whereas for dimensions outside this interval, there is none. 
This is consistent with the fact that there are no bound states in $d=2$ in the 
absence of a magnetic field in the critical regime \cite{mccoy}, although they might still be present deeper in the broken symmetry regime \cite{mussardo1,mussardo2,rychkov16}. Furthermore, our results show that no bound state is to be expected in dimension $d=4$. 

We have also studied the O$(2)$-symmetric model in $d=3$ along the same lines. Our results show the absence 
of a bound state in this case.

\section{Conclusions} 
\label{conclusion}
In this work we studied the existence of a bound state in the $\phi^4$ scalar theory in all dimensions between $d=2$ and $d=4$, and for a range of temperatures below the critical point. For $d=3$, our results are within $1\%$ of the previous Monte Carlo and numerical diagonalization values.  We use the BMW approximation of the Non-Perturbative Renormalization Group, which allows for the determination of the full-momentum dependence of the spectral function both in the universal and nonuniversal regimes. These results show once again the power of the BMW approximation for dealing with non-trivial physics at arbitrary momentum scales, even in cases where the quantities of interest require to perform analytic continuations of numerical data. 

Many generalizations of the present work can be envisaged. First, as the NPRG allows for the computation of nonuniversal quantities, studying
the presence of bound states for lattice models is a priori possible, since this only requires to take into account the lattice dispersion relation 
as was already done for the derivative expansion \cite{tristan,adam11}. Second, the dependence of the bound state spectrum on an external magnetic
field can be naturally studied within our formalism, since the $\rho$-dependence of  $\Gamma^{(2)}(p,\rho) $ encodes the influence of the external field on the spectral function. Finally, a BMW-type of approximation can also  be used to detect bound states in more 
complex systems, such as out-of-equilibrium \cite{kpz}, disordered \cite{tarjus, tissier} and quantum systems \cite{felix,adam14}, for which much 
less is known via simulations.

\begin{acknowledgments}
 We would like to thank N. Dupuis, A. Ran\c{c}on, J. Vidal and N. Wschebor for fruitful discussions. 
 F.B. wishes to thank the LPTMC for its hospitality during part of this work.
\end{acknowledgments}


\end{document}